\documentclass[12pt,onecolumn]{article}
\usepackage{amsfonts}
\usepackage{url}
\usepackage{amssymb}
\usepackage{multicol}
\usepackage{authblk}
\usepackage{graphicx}
\usepackage{float}
\usepackage{caption}
\usepackage{xcolor}
\usepackage[utf8]{inputenc}
\usepackage{amsmath}
\usepackage{hyperref}
\usepackage{nameref}

\makeatletter \@addtoreset{equation}{section}

\def \be{\begin{equation}}
\def \ee{\end{equation}}
\def \bea{\begin{eqnarray}}
\def \eea{\end{eqnarray}}

\newcommand{\nc}{\newcommand}
\nc{\al}{\alpha} \nc{\bib}{\bibitem} \nc{\la}{\lambda}
\nc{\C}{\mbox{\hspace{1.24mm}\rule{0.2mm}{2.5mm}\hspace{-2.7mm} C}}
\nc{\R}{\mbox{\hspace{.04mm}\rule{0.2mm}{2.8mm}\hspace{-1.5mm} R}}
\begin{document}

\title{\vspace{0cm}%
\rightline{\vspace
{0cm}} \textbf{Compactified 2HDM under the Non-SUSY AdS instability conjecture}\\}
\author[1,3]{M. A. Rbah%
\thanks{\href{mailto:Mohamedamin_rbah@um5.ac.ma}{Mohamedamin\_rbah@um5.ac.ma}}}
\author[1,3]{S. Saoud%
\thanks{\href{mailto:soulaimane_saoud@um5.ac.ma}{soulaimane\_saoud@um5.ac.ma}}}

\author[1,3]{R. Sammani%
\thanks{\href{mailto:rajae_sammani@um5.ac.ma}{rajae\_sammani@um5.ac.ma}}}

\author[1,2,3]{E. H. Saidi%
\thanks{\href{mailto:e.saidi@um5r.ac.ma}{
 e.saidi@um5r.ac.ma}}}
\author[1,3]{R. Ahl Laamara%
\thanks{\href{mailto:r.ahllaamara@um5r.ac.ma}{
 r.ahllaamara@um5r.ac.ma}}}

\affil[1]{\small LPHE-MS, Faculty of Sciences, Mohammed V University, Rabat, Morocco}

\affil[2]{\small Hassan II Academy of Science and Technology, Kingdom of Morocco}

\affil[3]{\small Centre of Physics and Mathematics, CPM-Morocco}
\maketitle

\begin{abstract}
We investigate how extra-dimensional dynamics influence Higgs-sector phenomenology
by compactifying a Two-Higgs-Doublet Model (2HDM) coupled to 4D gravity on a
circle \(S^1\). The resulting effective potential includes tree-level 2HDM
interactions, one-loop Coleman--Weinberg corrections from the Kaluza--Klein
towers, and an effective radion-stabilization contribution inspired by the
broader modulus-stabilization literature. We derive the corresponding 3D
effective action and show that, for the observed Higgs mass
\(m_h=125\,\mathrm{GeV}\), the compactified potential admits a stabilized
configuration with near-zero vacuum energy. By imposing the non-supersymmetric
AdS instability conjecture as a quantum-gravity consistency requirement, we
obtain a constraint within our numerical setup on the heavy Higgs sector,
finding that the additional scalar states must satisfy approximately
\(M_H \gtrsim 680\,\mathrm{GeV}\) in order to avoid perturbatively stable
non-supersymmetric AdS\(_3\) minima. Our results demonstrate how
Swampland-inspired constraints can yield phenomenologically relevant predictions
for extended Higgs sectors.

\end{abstract}


\section{Introduction}

\label{sec:Introduction} The discovery of a scalar boson with mass $%
125$ GeV at the LHC confirmed the Higgs mechanism as the origin of
electroweak symmetry breaking \cite{Aad,Chatrchyan_2012}. However,
remarkable as this achievement was, it did not resolve the long-standing
issues of naturalness nor clarify the broader landscape of
physics beyond the Standard Model (BSM) \cite{PDG2023,arhrib2021new}. Among
the simplest BSM extensions is the Two-Higgs-Doublet Model
(2HDM), which supplements the Standard Model by an additional $SU(2)_{L}$
doublet, and offers interesting phenomenology both in the scalar and flavor
sectors \cite{Branco2012,Ivanov2015,Hou:2024ibt} while remaining
economical in its parameter content. Yet, despite its appealing
structure, the 2HDM inherits the hierarchy problem and
motivates the exploration of mechanisms that can probe the stability and
consistency of extended scalar sectors beyond the purely four-dimensional
description.

Compactifying a four-dimensional theory on a circle $S^{1}$ provides
a calculable framework for investigating its lower-dimensional vacuum
structure. The resulting 3D effective field theory contains a radion field
which parametrizes the size of the compactified dimension and controls the
compactification radius \cite{Goldberger:1999uk,Goldberger:2000dv}. Quantum
effects from Kaluza--Klein towers, especially one-loop contributions, can
modify the effective potential and influence the existence and nature of
lower-dimensional extrema \cite{Coleman:1973jx,Das:2018radion}. In this
work, the radion is stabilized through an effective low-energy potential
ansatz, motivated by the broader radion-stabilization literature, rather
than through a unique microscopic derivation from a particular warped
compactification. Therefore, embedding the 2HDM in this framework provides
a controlled setting in which electroweak symmetry breaking, radion
stabilization, and compactification effects can be studied jointly.

Alongside these model-building considerations, Swampland
conjectures impose universal consistency conditions required of
any effective field theory to be compatible with quantum gravity. The
Weak Gravity Conjecture demands the existence of super-extremal
states to prevent the formation of stable black hole remnants \cite%
{ArkaniHamed:2007gg,Ooguri:2016pdq}, while the non-supersymmetric AdS
instability conjecture states that non-supersymmetric AdS vacua should not
be exactly stable in a consistent theory of quantum gravity \cite{Ooguri:2016pdq}.
These ideas have recently been applied to circle compactifications of the
Standard Model, yielding constraints on the masses of light particles
through quantum-vacuum-induced AdS$_{3}$ dynamics \cite%
{ArkaniHamed:2007gg,Ibanez:2017kvh,Castellano:2023}. Moreover, recent
progress in 3D gravity and higher-spin theories has sharpened our
understanding of the relevance of Swampland constraints to lower-dimensional
effective theories. This includes the demonstration of a finite 3D higher-spin
gravity landscape and the full classification of AdS$_3$ gauge sectors
\cite{sammani2024finiteness,sammani2023higher}, the construction of new
black-hole solutions in E6-extended 3D gravity \cite{sammani2025black},
recent progress on the asymptotic and minimal Weak Gravity Conjecture in
M-theory compactifications \cite{charkaoui2024asymptotic,charkaoui2025minimal},
and the formulation of the higher-spin swampland conjecture for massive
AdS$_3$ gravity \cite{sammani2025higher}.

In this work, we apply this ensemble of ideas to a compactified
2HDM coupled to four-dimensional gravity on a circle. This yields a
three-dimensional effective action involving radion--Higgs interactions,
one-loop Coleman--Weinberg corrections from scalar Kaluza--Klein modes,
and an effective radion stabilization potential. The latter is not intended
to represent a four-dimensional cosmological constant, nor is it claimed to
follow uniquely from the Goldberger--Wise mechanism. Rather, it is introduced
as a phenomenologically motivated stabilization ansatz for the radion sector,
with its own characteristic scale. We then subject this setup to Swampland
consistency conditions. First, we examine whether the measured Higgs
mass, $m_{h}=125$ GeV, admits a non-AdS$_{3}$ vacuum compatible
with both radion stabilization and electroweak symmetry breaking. Next,
we explore a range of heavy Higgs masses $M_{H}\in
\{400,600,680,780\}$ GeV and impose the non-SUSY AdS instability
criterion to identify regions of the parameter space that would lead to
perturbatively stable non-supersymmetric AdS$_3$ vacua. Within our numerical
setup, this criterion points to a lower bound of order
$M_{H}\gtrsim 680$ GeV. Our study thus illustrates how quantum-gravity
principles can constrain extended Higgs sectors and further affirms the
connection between extra-dimensional dynamics and low-energy phenomenology.

To close this introduction, we outline the structure of this
paper. In Section~\ref{sec:ModelSetup}, we construct the four-dimensional
2HDM coupled to gravity and carry out the circle compactification to
obtain the three-dimensional effective action. In Section~\ref{sec:EffPot},
we develop the full radion--Higgs potential where we take into account
the relevant interactions, scalar one-loop KK corrections, and a
phenomenologically motivated radion stabilization ansatz. In
Section~\ref{sec:Swamp}, we introduce the relevant Swampland conjectures
and formulate the precise consistency checks that our compactified model
will be subject to. In Section~\ref{sec:PhenImp}, we apply these
quantum-gravity consistency conditions to both the measured Higgs mass of
125 GeV, as well as to a scan of heavier Higgs masses. Finally,
Section~\ref{sec:Discussion} discusses the resulting implications for
Higgs-sector phenomenology and highlights promising directions for future
investigations.

\section{Model Setup}
\label{sec:ModelSetup}
This section outlines the theoretical framework for our study. We begin with a review of the scalar potential and parameter space of the Two-Higgs-Doublet Model (2HDM) with experimental limits imposed, and then we embed the 2HDM as a four-dimensional theory of gravity, which we will utilize as the fundamental underpinnings for our analysis of radion-Higgs dynamics later.
\subsection{The Two-Higgs-Doublet Model Framework}
The Two-Higgs-Doublet Model (2HDM) builds on the Standard Model by adding an additional scalar doublet. The Higgs sector is enriched, giving rise to new sources of CP violation, possible dark matter candidates, and new Yukawa structures. To guarantee that the model does not allow for flavor-changing neutral currents, a discrete Z2 symmetry is imposed under which the two doublets behave differently. When this symmetry is softly broken, the scalar potential is still renormalizable and phenomenologically viable, and respects the richness of the dynamics of the extended Higgs sector.
Under a softly broken discrete \(Z_2\) symmetry (\(\Phi_1\to+\Phi_1,\;\Phi_2\to-\Phi_2\)), the most general CP-conserving, \(SU(2)_L\times U(1)_Y\) invariant scalar potential of the 2HDM is given by \cite{Branco2012,Ivanov2015}
\begin{equation}
\begin{aligned}
V_{tree} &=\,m_{11}^2\,\Phi_1^\dagger\Phi_1 + m_{22}^2\,\Phi_2^\dagger\Phi_2 
   - m_{12}^2\bigl(\Phi_1^\dagger\Phi_2+\text{h.c.}\bigr)\\
&\quad +\tfrac{\lambda_1}{2}\,(\Phi_1^\dagger\Phi_1)^2
   +\tfrac{\lambda_2}{2}\,(\Phi_2^\dagger\Phi_2)^2
   +\lambda_3\,(\Phi_1^\dagger\Phi_1)(\Phi_2^\dagger\Phi_2)\\
&\quad +\lambda_4\,(\Phi_1^\dagger\Phi_2)(\Phi_2^\dagger\Phi_1)
   +\tfrac{\lambda_5}{2}\,\bigl[(\Phi_1^\dagger\Phi_2)^2+\text{h.c.}\bigr],
\end{aligned}
\end{equation}
where all \(m_{ij}^2\) and \(\lambda_i\) are real parameters, and terms proportional to \(\lambda_{6,7}\) are absent at tree level by the \(Z_2\) symmetry \cite{Branco2012}.

Each doublet can be decomposed as \cite{Haber1985,Davidson2005}  
\begin{equation}
\Phi_i = 
\begin{pmatrix}
\phi_i^+\\
\dfrac{v_i + \rho_i + i\,\eta_i}{\sqrt{2}}
\end{pmatrix},
\quad i=1,2,
\end{equation}
with vacuum expectation values \(v_i\), charged fields \(\phi_i^+\), CP-even scalars \(\rho_i\), and CP-odd scalars \(\eta_i\). Electroweak symmetry breaking enforces \(v\equiv\sqrt{v_1^2+v_2^2}=246\)\,GeV.

The real parameters \(\{m_{11}^2,m_{22}^2,m_{12}^2,\lambda_{1--5}\}\) are reduced by the two minimization conditions of \(V_{2\mathrm{HDM}}\), trading \(m_{11}^2\) and \(m_{22}^2\) for \(\{v_1,v_2\}\). The remaining seven physical parameters may be chosen as \(\{M_h,M_H,M_A,M_{H^\pm},\alpha,\tan\beta,m_{12}^2\}\), where \(\alpha\) is the CP-even mixing angle and \(\tan\beta= v_2/v_1\) \cite{Ivanov2015,Ferreira2015}.

Experimentally, the observed SM-like Higgs boson mass is  
\begin{equation}
  m_h \simeq (125.10\pm0.14)\,\mathrm{GeV},
\end{equation}
as compiled by the Particle Data Group \cite{PDG2023}.  

Flavor physics, specifically the rare decay $B\to X_s\gamma$, provides a rich source of indirect bounds on the charged Higgs mass, being very sensitive to charged Higgs contributions at loop level. Indirect bounds are modified in the presence of vector-like quarks (VLQs), because the additional contributions to the effective operators that describe this process. Current analyses show that 
\[
   m_{H^\pm} \;\gtrsim\; 400~\text{GeV (singlet VLQ)}, 
   \qquad 
   m_{H^\pm} \;\gtrsim\; 450~\text{GeV (doublet VLQ)},
\]
under 95\% CL in 2HDM-II+VLQ scenarios\cite{Benbrik2024}.
These results show that, despite still being theoretically allowed in minimal 2HDM setups, low charged-Higgs masses are very strongly constrained when VLQ effects are included.
Complementary information comes from direct collider searches, via di-Higgs production in the alignment limit, which is where the observed 125 GeV Higgs boson appears to behave Standard Model-like, and the additional CP-even or CP-odd scalar states contribute substantially to the di-Higgs production cross section. Searches at the LHC show that new scalar states associated with top-Yukawa-enhanced scenarios with mass below $\sim 300$–$400$ GeV are strongly disfavored\cite{Iguro2023}. Accordingly, the collider bounds reinforce the flavor bounds and effectively move the accessible heavy Higgs spectrum to several hundred GeV.

\bigskip

Having reviewed the scalar sector of the standard 2HDM, we now embed it within a
four-dimensional theory with dynamical gravity. This step is not intended to
determine the ultraviolet properties of the Higgs sector, but to specify
the quantum effective field theory that is going to be compactified on a circle of radius.

In this setup, the radion is a scalar degree of freedom
As for the size of the compact direction and its coupling to Higgs

This sector gives us an opportunity to analyze radion--Higgs dynamics in a lower-dimensional

effective theory.

We start with the Einstein--Hilbert action supplemented by a

moduli vacuum-energy term in four-dimension and the scalar sector of 2HDM. The

total four-dimensional action is
\begin{equation}
\label{eq:4D_action}
S_{4D}
= \int d^4x\,\sqrt{-g_{(4)}}\,
\biggl[
\frac{M_4^2}{2}\,\mathcal{R}_{(4)} - \Lambda_4
+ \sum_{i=1}^2 (D_M\Phi_i)^\dagger (D^M\Phi_i)
- V_{tree}(\Phi_1,\Phi_2)
\biggr],
\end{equation}
where \(M_4\) denotes the four-dimensional Planck mass,
\(\mathcal{R}_{(4)}\) is the four-dimensional Ricci scalar, and
\(\Lambda_4\) is a four-dimensional vacuum-energy parameter. We emphasize
that \(\Lambda_4\) is distinct from the radion-stabilization scale
\(\Lambda_{\rm rad}\) introduced later in the effective radion potential.

The covariant derivative acting on each scalar doublet is defined as
\begin{equation}
D_M = \partial_M 
- i\,g\,W_M^a T^a 
- i\,g'\,Y_\Phi B_M ,
\end{equation}
where \(T^a = \sigma^a/2\) are the generators of \(SU(2)_L\),
\(Y_\Phi=1/2\) is the hypercharge of the Higgs doublets, and
\(M=0,1,2,3\). The scalar fields \(\Phi_1\) and \(\Phi_2\) therefore
transform as weak doublets with hypercharge \(Y_\Phi=+1/2\), and under a
local gauge transformation they obey
\begin{equation}
\Phi_i(x) \;\to\; 
e^{i\alpha^a(x)T^a + i\beta(x)Y_\Phi}\,\Phi_i(x).
\end{equation}

The scalar sector Lagrangian thus takes the compact form:
\begin{equation}
\mathcal{L}_{\text{scalar}} = 
  \sum_{i=1}^2 (D_M\Phi_i)^\dagger (D^M\Phi_i)
  - V_{tree}(\Phi_1,\Phi_2),
\end{equation}
which is directly included in the total action \eqref{eq:4D_action}. The structure encapsulates both gravitational and electroweak scalar dynamics in a consistent four-dimensional field-theoretic framework.

\subsection{Circle Compactification of the 2HDM+GR Action}
\paragraph{}

Compactifying four-dimensional field theories on a circle offers a natural
framework to study the vacuum structure of lower-dimensional effective
theories. In particular, the effective potential arising from compactification,
\(V_{\text{eff}}\), captures the fluctuations of the radion and the quantum
corrections induced by Kaluza--Klein modes, together with the contribution
inherited from the four-dimensional vacuum energy. This potential determines
the stability of the compactified geometry. For lower-dimensional scalar
sectors, such as that of the 2HDM, compactification has phenomenological
relevance. The radion couples to the Higgs fields, thereby modifying their
potential and possibly shifting their vacuum expectation values or their mass
spectrum. The radion--Higgs potential can have non-trivial extrema that lead to
effective three-dimensional anti-de Sitter (AdS), Minkowski, or de Sitter (dS)
vacua. In addition to enlarging the vacuum structure, such systems provide a
useful test case for quantum-gravity conjectures in a phenomenologically
motivated Higgs sector.

\paragraph{}
Importantly, similar techniques have been applied to the Standard Model (SM)
itself. In \cite{ArkaniHamed:2007gg}, it was shown that the SM exhibits a
landscape of lower-dimensional vacua arising from vacuum and Casimir energies
when compactified on a circle. Moreover, in \cite{Ibanez:2017kvh}, the WGC was
used to derive limits on neutrino masses and the cosmological constant, thereby
connecting particle-physics parameters with quantum-gravity constraints.
Furthermore, \cite{Castellano:2023} used similar circle compactification
techniques to investigate how towers of states and hierarchical structures in
the SM can emerge from quantum-gravity considerations. These works demonstrate
that circle compactification can generate lower-dimensional AdS extrema and
therefore provide a sensitive probe of Swampland constraints.

Motivated by these results, it is reasonable to ask how an extended Higgs sector
behaves under circle compactification. The Two--Higgs--Doublet Model is
a particularly interesting case, because the radion couples not only to the
SM-like Higgs mode but also to the additional scalar degrees of freedom. We
therefore consider the 2HDM minimally coupled to four-dimensional gravity, with
the action given in \eqref{eq:4D_action}.

We compactify one spatial direction on a circle \(S^1\) with background radius
\(R\), introducing the coordinate \(y\in[0,2\pi)\). In the four-dimensional
Einstein frame, we use the metric ansatz
\begin{equation}
ds_{(4)}^2
=
\frac{r^2}{R_{\rm phys}^2(\phi)}\,ds_{(3)}^2
+
R_{\rm phys}^2(\phi)
\left(
dy-\frac{\sqrt{2}}{M_4 r}V_\mu(x)\,dx^\mu
\right)^2 .
\label{eq:metric_ansatz}
\end{equation}
Here, the graviphoton \(V_\mu\) arises from the off-diagonal component of the
Einstein-frame metric tensor \(g_{\mu y}\). The parameter \(r\), defined by the
Einstein-frame parametrization, is a fixed reference length, while the
compactification radius \(R\) sets the size of the compactified circle, whose
volume is \(2\pi R\). The radion-field normalization scale \(R_0\) will be
introduced below.

The physical radius of the compactified circle is defined in terms of the
radion field as
\begin{equation}
R_{\rm phys}(\phi)
=
R\,\exp\left(\frac{\phi}{R_0}\right),
\label{eq:Rphys}
\end{equation}
where \(R_0\) is introduced to make the ratio \(\phi/R_0\) dimensionless. For
\(\phi=0\), one obtains
\begin{equation}
R_{\rm phys}(0)=R,
\end{equation}
So \(R\) is the unperturbed radius of the compactification. This notation
distinguishes \(R_{\rm phys}\), the effective geometrical scale of the
compactification, from \(\phi\), the field coordinate of the radion field.

The vacuum structure can be equivalently analyzed in terms of \(\phi\), since
\(R_{\rm phys}(\phi)\) is a monotonic and invertible function.

As for the compactification formulas, they retain their standard geometrical
form in terms of \(R_{\rm phys}\) for use in the analysis of the vacuum
structure.

Substituting the ansatz into the Einstein--Hilbert term and integrating over
the compact coordinate \(y\), one obtains the gravitational contribution to the
three-dimensional effective action:
\begin{equation}
\begin{aligned}
S_{\mathrm{grav}}
&=
\int d^3x\,dy\,\sqrt{-g_{(4)}}\,
\left[
\tfrac12 M_4^2\,\mathcal{R}_{(4)}
-
\Lambda_4
\right]
\\
&\longrightarrow
\int d^3x\,\sqrt{-g_{(3)}}\,(2\pi R)\,
\Biggl[
\tfrac12 M_4^2\,\mathcal{R}_{(3)}
-
\tfrac14
\frac{R_{\rm phys}^4(\phi)}{r^4}
F_{\mu\nu}F^{\mu\nu}
-
M_4^2
\bigl(\partial_\mu\ln R_{\rm phys}(\phi)\bigr)^2
-
\frac{r^2\Lambda_4}{R_{\rm phys}^2(\phi)}
\Biggr],
\end{aligned}
\label{eq:grav_reduction}
\end{equation}
where
\begin{equation}
F_{\mu\nu}
=
\partial_\mu V_\nu-\partial_\nu V_\mu .
\end{equation}
The last term in \eqref{eq:grav_reduction} is inherited from the
four-dimensional vacuum-energy contribution. It should not be confused with
the independent radion-stabilization contribution introduced later in the
effective potential.

Using \eqref{eq:Rphys}, the radion kinetic term may also be written as
\begin{equation}
\bigl(\partial_\mu\ln R_{\rm phys}(\phi)\bigr)^2
=
\frac{1}{R_0^2}\,
\partial_\mu\phi\,\partial^\mu\phi .
\end{equation}
Thus, one may work either in the geometrical radius variable
\(R_{\rm phys}\) or in the radion field coordinate \(\phi\). In the vacuum
analysis below, we use \(\phi\) as the dynamical radion variable.

The Higgs doublets \(\Phi_i(x,y)\) admit a Kaluza--Klein expansion along the
circle:
\begin{equation}
\Phi_i(x,y)
=
\frac{1}{\sqrt{2\pi R}}
\sum_{n=-\infty}^{+\infty}
\Phi_i^{(n)}(x)\,e^{iny},
\qquad i=1,2 .
\end{equation}
At energies below the compactification scale, \(E\ll R^{-1}\), only the zero
modes remain light in the low-energy effective theory. The massive
Kaluza--Klein modes are then integrated out, and their leading scalar-sector
effects are included through the one-loop Coleman--Weinberg correction. Since
the physical KK scale is controlled by the radion through
\begin{equation}
M_{\rm KK}(\phi)
\sim
\frac{1}{R_{\rm phys}(\phi)}
=
\frac{1}{R}\exp\left(-\frac{\phi}{R_0}\right),
\end{equation}
the one-loop contribution is naturally radion dependent.

After integrating over the compact circle, the effective three-dimensional
action takes the form
\begin{equation}
\begin{aligned}
S_{3\mathrm{D}}
=
\int d^3x\,\sqrt{-g_{(3)}}\,
\Biggl[
&2\pi R
\left(
\tfrac12 M_4^2\,\mathcal{R}_{(3)}
-
\tfrac14
\frac{R_{\rm phys}^4(\phi)}{r^4}
F_{\mu\nu}F^{\mu\nu}
-
\frac{r^2\Lambda_4}{R_{\rm phys}^2(\phi)}
\right)
\\
&-
(2\pi R)M_4^2
\bigl(\partial_\mu\ln R_{\rm phys}(\phi)\bigr)^2
+
2\pi R\,\mathcal{L}_{\rm matter}
\Biggr].
\end{aligned}
\label{eq:S3D_first}
\end{equation}
Defining the three-dimensional Planck mass by
\begin{equation}
M_3^2
=
2\pi R\,M_4^2 ,
\end{equation}
this can be rewritten as
\begin{equation}
\begin{aligned}
S_{3\mathrm{D}}
=
\int d^3x\,\sqrt{-g_{(3)}}\,
\Biggl[
&
\tfrac12 M_3^2\,\mathcal{R}_{(3)}
-
\frac{\pi R}{2}
\frac{R_{\rm phys}^4(\phi)}{r^4}
F_{\mu\nu}F^{\mu\nu}
-
M_3^2
\bigl(\partial_\mu\ln R_{\rm phys}(\phi)\bigr)^2
\\
&+
2\pi R\,\mathcal{L}_{\rm matter}
-
2\pi R
\frac{r^2\Lambda_4}{R_{\rm phys}^2(\phi)}
\Biggr].
\end{aligned}
\label{eq:S3D_second}
\end{equation}

The matter sector is written as
\begin{equation}
\mathcal{L}_{\rm matter}
=
\sum_{i=1}^{2}
(D_\mu\Phi_i)^\dagger(D^\mu\Phi_i)
-
\left(
V_{\rm tree}^{3D}(\Phi_i)
+
V_{\rm rad}(\phi)
+
V_{\rm int}(\phi,\Phi_i)
+
V_{\rm loop}^{(1)}(\phi,\Phi_i)
\right).
\end{equation}
Here \(V_{\rm tree}^{3D}\) is the compactified tree-level 2HDM scalar
potential, \(V_{\rm rad}\) is an effective radion-stabilization potential,
\(V_{\rm int}\) describes the radion--Higgs interaction, and
\(V_{\rm loop}^{(1)}\) denotes the scalar one-loop correction induced by the
Kaluza--Klein tower.

Thus, the final expression of the three-dimensional action is
\begin{equation}
\begin{aligned}
S_{3\mathrm{D}}
&=
\int d^3x\,\sqrt{-g_{(3)}}\,
\Biggl[
\tfrac12 M_3^2\,\mathcal{R}_{(3)}
-
\frac{\pi R}{2}
\frac{R_{\rm phys}^4(\phi)}{r^4}
F_{\mu\nu}F^{\mu\nu}
-
M_3^2
\bigl(\partial_\mu\ln R_{\rm phys}(\phi)\bigr)^2
\\
&\quad
+
2\pi R
\left(
\sum_{i=1}^{2}
(D_\mu\Phi_i)^\dagger(D^\mu\Phi_i)
-
V_{\rm eff}^{(3D)}(\phi,\Phi_i)
\right)
\Biggr].
\end{aligned}
\label{eq:3D_action_final}
\end{equation}
These correspond to a geometric definition of the variable
\(R_{\rm phys}(\phi)\), coming from dimensional reduction, namely from the
metric and its dynamical degrees of freedom in Kaluza--Klein theory. Instead,
one can take \(\phi\) as the scalar variable in the effective potential of the
scalar sector and consider this argument as a vacuum coordinate in the effective
potential \(V_{\rm eff}^{(3D)}(\phi,\Phi_i)\). With the relation
\(R_{\rm phys}(\phi)=R\exp(\phi/R_0)\), we can define how the gravitational
length behaves depending on \(\phi\). The terms
\(V_{\rm rad}(\phi)\), \(V_{\rm int}(\phi,\Phi_i)\), and
\(V_{\rm loop}^{(1)}(\phi,\Phi_i)\) contribute to the scalar spectrum depending
on \(R\).

The effective potential is therefore
\begin{equation}
V_{\rm eff}^{(3D)}(\phi,\Phi_i)
=
V_{\rm tree}^{3D}(\Phi_i)
+
V_{\rm rad}(\phi)
+
V_{\rm int}(\phi,\Phi_i)
+
V_{\rm loop}^{(1)}(\phi,\Phi_i)
+
\frac{r^2\Lambda_4}{R_{\rm phys}^2(\phi)} .
\label{eq:Veff_3D_general}
\end{equation}
The final term in \eqref{eq:Veff_3D_general} comes from the original
four-dimensional vacuum-energy contribution after dimensional reduction. By
contrast, \(V_{\rm rad}(\phi)\) is an independent effective stabilization
ansatz for the radion sector.

In the following section, we give a precise definition of the full
three-dimensional effective potential, including its radion-dependent
structure, the radion--Higgs interaction, and the scalar Kaluza--Klein quantum
corrections.

\section{Effective potential from 2HDM Compactification}
\label{sec:EffPot}
\paragraph{}
In this section, we analyze the three-dimensional effective potential associated
with the compactification of the Two-Higgs-Doublet Model with gravity on a
circle. First, we discuss the tree-level structure obtained from the
Kaluza--Klein expansion of the Higgs fields, then we add one-loop quantum
corrections which may change the vacuum structure. The compact dimension is
stabilized by an effective radion potential, and radion--Higgs interaction
terms are added in order to encode the dependence of the Higgs sector on the
size of the extra dimension. \\

When we compactify a higher-dimensional theory such as the
Two-Higgs-Doublet Model with gravity down to three dimensions, new fields
show up in the lower-dimensional effective theory. One important field is the
dilatonic field, namely the radion \(\phi\), which describes fluctuations of
the physical size of the extra dimension. Therefore, the effective potential
depends on the radion field as well as on the Higgs fields. There are several
elements in this potential: the tree-level contribution inherited from the
original four-dimensional scalar potential, an effective stabilization
potential for the radion, non-linear interaction terms between the radion and
the Higgs fields, and quantum corrections coming from the Kaluza--Klein tower.

\subsection{One-Loop Structure of the 2HDM Scalar Potential}
\paragraph{}
The Higgs fields \( \Phi_1 \) and \( \Phi_2 \) are scalar fields that can be expanded in terms of their Kaluza-Klein modes as follows

\[
\Phi_1 (x, y) = \frac{1}{\sqrt{2\pi R}} \sum_n \Phi_1^{(n)}(x^\mu) e^{iny / R}.
\]
\[
\Phi_2 (x, y) = \frac{1}{\sqrt{2\pi R}} \sum_n \Phi_2^{(n)}(x^\mu) e^{iny / R}.
\]

where \( \Phi_1^{(n)}(x^\mu) \) and \( \Phi_2^{(n)}(x^\mu) \) are the field components of the Higgs doublets in 3D, and \( y \) is the coordinate of the extra dimension, compactified on a circle of radius \( R \).

At tree level, for each KK mode, the potential in 3D becomes a sum over the KK modes as follows

\begin{align}
V_{\rm tree}^{3D}=2\pi RV_{\rm tree} ,
\end{align}
with 
\begin{align}
   V_{\rm tree}  &= \frac{1}{2\pi R}\sum_{n} \Bigl[ 
   \bigl(m_{11}^2 + \frac{n^2}{R^2}\bigr)\, \Phi_1^{(n)\dagger} \Phi_1^{(n)}
 + \bigl(m_{22}^2 + \frac{n^2}{R^2}\bigr)\, \Phi_2^{(n)\dagger} \Phi_2^{(n)}
 - m_{12}^2 \bigl( \Phi_1^{(n)\dagger} \Phi_2^{(n)} + \Phi_2^{(n)\dagger} \Phi_1^{(n)} \bigr)
\Bigr] \nonumber\\
&\quad + \frac{1}{2\pi R} \sum_{n} \Biggl[
 \frac{\lambda_1}{2} (\Phi_1^{(n)\dagger} \Phi_1^{(n)})^2
+ \frac{\lambda_2}{2} (\Phi_2^{(n)\dagger} \Phi_2^{(n)})^2
+ \lambda_3 (\Phi_1^{(n)\dagger} \Phi_1^{(n)}) (\Phi_2^{(n)\dagger} \Phi_2^{(n)}) \nonumber\\
&\qquad + \lambda_4 |\Phi_1^{(n)\dagger} \Phi_2^{(n)}|^2
+ \frac{\lambda_5}{2} \Bigl[ (\Phi_1^{(n)\dagger} \Phi_2^{(n)})^2 + \text{h.c.} \Bigr]
\Biggr].
\label{eq:Vtree3D}
\end{align}
In the potential expressions, $n$ labels the Kaluza-Klein modes corresponding to discrete momentum values from the compactified dimension, and $R$ is the compactification radius.
The compactified extra dimension, physically, changes the vacuum energy by the KK tower of quantum states. The contribution from large levels n in the KK spectrum, however, is exponentially suppressed as $\propto e^{-2\pi n}$ and illustrates the decoupling of heavy modes from low energy 3D dynamics. These excitations may slightly distort the shape of the scalar potential, shift vacuum expectation values, or introduce new local minima, depending on the stabilization scale and couplings for the model\cite{Ibanez:2017kvh}.

\paragraph{}
Radiative corrections can modify the shape of the potential, shift the position
of the vacuum, generate additional local minima, and change the stability of the
scalar-field configuration \cite{Coleman:1973jx}. In the Two-Higgs-Doublet
Model, it has been shown that one-loop effects can produce vacua that
spontaneously break charge conservation and that are absent at tree level,
thereby showing the importance of including the Coleman--Weinberg contribution
\cite{Ferreira:2019wjz}. Recent investigations have also extended this
framework by considering global symmetries in orbit space and thermal
corrections, confirming that one-loop corrections can modify the symmetries
preserved at tree level and alter the phase structure of the model
\cite{Cao:2023npo,Castellano:2023}.

At one loop, the quantum correction to the effective potential can be written
in the Coleman--Weinberg form
\begin{equation}
\label{eq:VCWgeneral}
V_{\rm CW}
=
\sum_i
\frac{n_i\,M_i^4}{64\pi^2}
\left[
\ln\!\left(\frac{M_i^2}{\mu^2}\right)
-
C_i
\right],
\end{equation}
In this notation, the index \(i\) labels the fluctuating degrees of freedom,
while \(n_i\) indicates their multiplicity. For bosonic degrees of freedom,
\(n_i\) is positive, whereas for fermionic degrees of freedom, \(n_i\) is
negative. The parameter \(\mu\) is the renormalization scale, and \(C_i\) is a
scheme-dependent constant. For example, in the \(\overline{\rm MS}\) scheme,
one has \(C_i=3/2\) for scalar degrees of freedom
\cite{Coleman:1973jx}.

In the present investigation, we consider only the contribution from the scalar
fields of the compactified 2HDM to the scalar Kaluza--Klein
Coleman--Weinberg potential. Therefore, the scalar KK contribution should be
regarded as the compactified 2HDM scalar-sector contribution to the total
one-loop effective potential; that is, it does not include contributions from
gauge bosons, fermions, or ghosts. The scalar KK tower is then described as Eq.~\eqref{eq:VCWgeneral} gives
\begin{equation}
\label{eq:Vloop}
V_{\rm loop}^{\rm KK}(R,\Phi_i)
=
\sum_{n}
\frac{M_{\rm eff,n}^4(\Phi_i)}{64\pi^2}
\left[
\ln\!\left(
\frac{M_{\rm eff,n}^2(\Phi_i)}{\mu^2}
\right)
-\frac{3}{2}
\right],
\end{equation}
where the field-dependent mass of the \(n\)-th Kaluza--Klein scalar mode is
\begin{equation}
M_{\rm eff,n}^2(\Phi_i)
=
M^2(\Phi_i)
+
\frac{n^2}{R^2}.
\end{equation}
The term \(M_i^2(\Phi)\) denotes a field-dependent scalar mass eigenvalue
arising from the scalar sector of the Two-Higgs-Doublet Model (2HDM). The
Kaluza--Klein mass spectrum is fixed by the background compactification radius
\(R\).

The infinite sum over Kaluza--Klein modes appearing in Eq.~\eqref{eq:Vloop} is
not treated as an unregularized divergent quantity. As in the standard
Coleman--Weinberg framework, the ultraviolet-divergent pieces are absorbed into
the renormalized parameters and counterterms of the effective theory. After this
renormalization procedure, the physically relevant finite-radius contribution is
well defined. Equivalently, using a Poisson-resummed representation of the
Kaluza--Klein tower, the radius-dependent part can be expressed as a sum
involving modified Bessel functions, whose arguments depend on the
compactification radius.

For massive Kaluza--Klein states, the large-mode behavior is exponentially
suppressed.
\begin{equation}
K_{\nu}\!\left(2\pi n M R\right)
\sim
\sqrt{\frac{1}{4 n M R}}\,
e^{-2\pi n M R},
\qquad n\gg 1 .
\end{equation}
The one-loop effective potential has what is known as the standard decoupling
of massive compactified modes, resulting in exponential behavior for the
finite-radius part of the potential \cite{Ponton:2001hq}. The effects of
massive Kaluza--Klein modes on the low-energy three-dimensional theory provide
a controlled source of radiative corrections and are expected to make a small
contribution at low energy, though they can induce finite shifts in the scalar
potential and may change the location and/or stability of local extrema.

As discussed in the previous sections, the truncation of the Casimir
contribution remains an open question. Contributions from massless or
ultra-light fields have not been assessed in this analysis. The properties of
the complete light spectrum in the compactified theory include contributions
from light fields, including their spin, particle statistics, and the boundary
conditions under which fields propagate around the compactified circle
\cite{Ponton:2001hq}. Massless fields have the potential to generate
power-law radius-dependent terms, unlike massive modes, which generate
exponentially suppressed terms, and may therefore play a significant role in
determining the radion dynamics. The mechanism behind the sensitivity of
lower-dimensional Standard Model compactifications to massless fields is
associated particularly with neutrinos
\cite{Ibanez:2017kvh,inspirehep1926195}. Including light fields, specifically
massless and ultra-light fields, would require specifying the full particle
content and the corresponding boundary conditions; therefore, additional
light-field Casimir contributions are not included in the present effective
truncation.

\subsection{Radion Potential and Higgs Interaction}

In order to stabilize the compact direction, we lift the radius modulus.
Thus, an effective potential for the radion admitting a local minimum is added
to the compactified 2HDM scalar sector. In the present analysis, this
stabilization sector is modeled by
\begin{equation}
V_{\rm rad}(\phi)
=
\Lambda_{\rm rad}
\left[
1-\exp\left(-\frac{\phi}{R_0}\right)
\right]^2 .
\label{eq:Vrad}
\end{equation}
The quantity \(R_0\) serves as a fixed reference length for the scale of the
extra-dimensional geometry, so that \(\phi/R_0\) is dimensionless, while
\(\Lambda_{\rm rad}\) controls the overall scale of the radion-stabilization
sector in the three-dimensional effective theory.

The potential for the radion field, \(V_{\rm rad}\), is used as a low-energy
parametrization of radion stabilization. It is not tied to any particular
microscopic Randall--Sundrum realization, but is motivated by the general fact
that extra-dimensional moduli can acquire potentials through bulk dynamics,
quantum effects, or other stabilization mechanisms
\cite{Goldberger:1999uk,Goldberger:2000dv,Das:2018radion}. Accordingly,
\(\Lambda_{\rm rad}\) denotes the scale associated with the radion sector and
should not be identified with a four-dimensional cosmological constant.

The radion potential possesses a stationary point corresponding to the
reference value \(\phi=0\). Indeed,
\begin{equation}
\left.
\frac{dV_{\rm rad}}{d\phi}
\right|_{\phi=0}
=0,
\qquad
\left.
\frac{d^2V_{\rm rad}}{d\phi^2}
\right|_{\phi=0}
=
\frac{2\Lambda_{\rm rad}}{R_0^2}.
\end{equation}
Thus, for \(\Lambda_{\rm rad}>0\), the radion direction has positive curvature
around \(\phi=0\), so that the compactification radius is locally stabilized.

\paragraph{}
Even after the radius modulus becomes stable, there will still be fluctuations
in the radius modulus which couple to the Higgs sector. The effect of this
coupling can be seen at low energies through radion--Higgs operators,
suppressed by a certain interaction scale. The leading term is
\(V_{\rm int}(\phi,\Phi_i)\).

The expression for the interaction term is as follows:
\begin{equation}
V_{\rm int}(\phi,\Phi_i)
=
\frac{\phi}{\Lambda_\phi R_0^2}
\left(
\lambda_{\phi1}|\Phi_1|^2
+
\lambda_{\phi2}|\Phi_2|^2
\right)^2 ,
\end{equation}
where \(\lambda_{\phi1}\) and \(\lambda_{\phi2}\) are the two constants used to
describe the radion coupling with both Higgs doublet fields.

The operator described above is made up of gauge-invariant Higgs bilinears
\(|\Phi_1|^2\) and \(|\Phi_2|^2\), and represents the leading effective
coupling between the radion and the Higgs sector. These types of couplings naturally arise when the Higgs sector couples to
either the compactification scale or the metric-modulus fluctuations
\cite{Giudice:2000av,Barger:2012qva,Eroncel:2020et}. In this effective
theory, the parameters \(\Lambda_\phi\), \(\lambda_{\phi1}\), and
\(\lambda_{\phi2}\) are free parameters of the compactified theory, and they
represent how radion fluctuations modify the Higgs potential and subsequently
how they modify the vacuum structure of the three-dimensional model.

\section{Swampland constraints on Radion-induced Higgs phenomenology }
The presence of a stabilized radion field in compactified configurations has remarkable
implications for low-energy phenomenology. Particularly, the radion field couples to the
Higgs sector in such a way that modifies the effective potential and thus alters electroweak
symmetry breaking and the resulting mass spectrum. When the radion field is coupled to the
Higgs doublets, additional contributions arise in the total effective potential that explicitly
depend on the compactification radius. These contributions can stabilize a vacuum
state in which the physical size of the extra dimension and the values of fundamental
parameters, such as the Higgs mass, are dynamically correlated.
The framework developed here therefore establishes a predictive relation between the
radion stabilization parameter and the measured Higgs mass. Below, we present our
numerically evaluated examples of the effective potential for a number of inputs to the
Higgs mass.

\label{sec:Swamp}
\subsection{Radion induced phenomenology in the extended Higgs sector}
\label{sec:PhenImp}
To demonstrate how the compactification dynamics modifies the vacuum structure of the theory, we investigate the behaviour of the radion-dependent effective potential \(V_{\text{eff}}(R)\) for two representative choices of Higgs-sector masses. In this subsection, we analyse only the phenomenological implications of these configurations.

\medskip

\begin{figure}[H]
    \centering
    \includegraphics[width=0.75\linewidth]{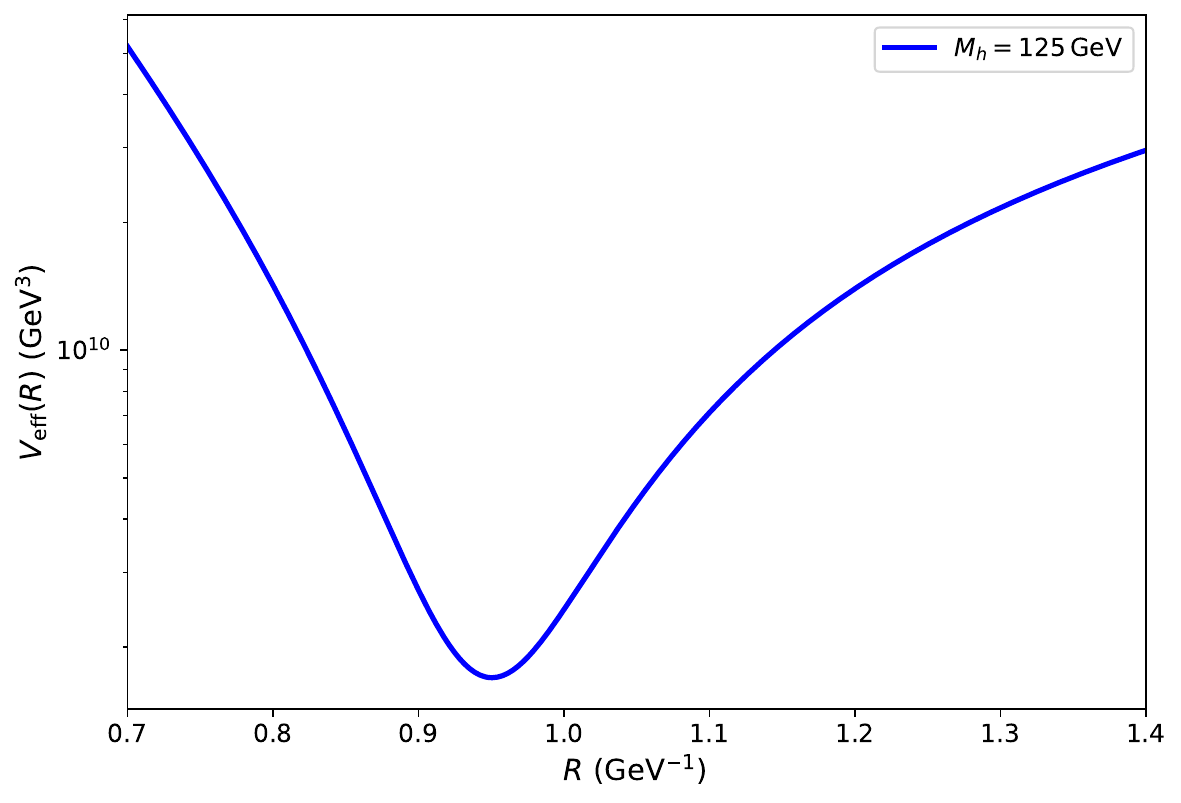}
    \caption{Effective potential \(V_{\text{eff}}(R)\) for a light Higgs mass \(M_h = 125\,\mathrm{GeV}\), evaluated at \(\mu = 240\,\mathrm{GeV}\). The minimum occurs at positive vacuum energy.}
    \label{fig:lightH}
\end{figure}

Figure~\ref{fig:lightH} illustrates the effective potential as a function of a light Higgs mass input \(M_h = 125\,\mathrm{GeV}\) at the renormalization scale \(\mu = 240\,\mathrm{GeV}\), $\Lambda = 1 \times 10^{12}\,\mathrm{GeV}^4  $ and $\ \phi = 10^{10}\,\mathrm{GeV}^{-1}.$
The potential displays a smooth and well-defined minimum. Importantly, the effective potential is positive at the minimum. This corresponds to a configuration that is phenomenologically safe in the context of the radion: the radion is stabilized around the compactification, the compactification is consistent with non-negative vacuum energy density, and the contribution from the light Higgs sector only contributes mild curvature to the effective potential. The shallow profile of the effective potential reflects that the potential is only mildly influenced by the light scalar mass.

\begin{figure}[H]
    \centering
    \includegraphics[width=0.75\linewidth]{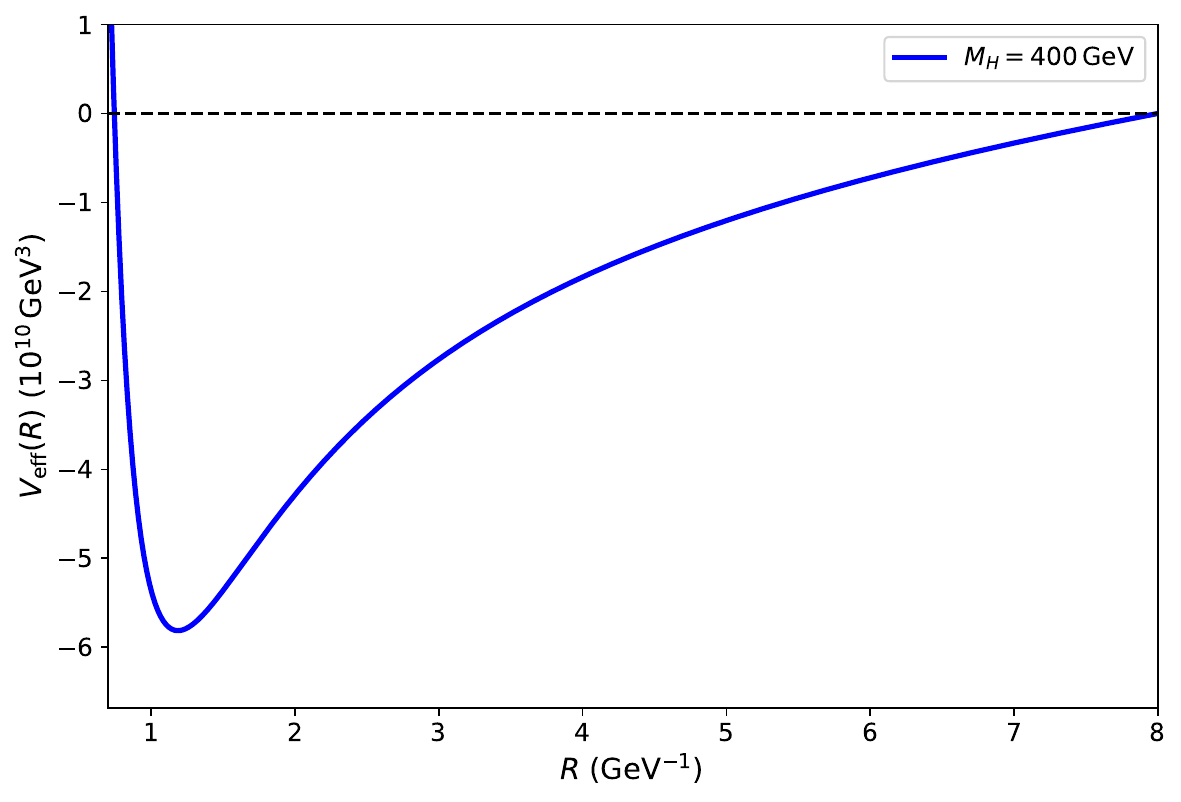}
    \caption{Effective potential \(V_{\text{eff}}(R)\) for a heavy Higgs mass \(M_H = 400\,\mathrm{GeV}\), evaluated at \(\mu = 340\,\mathrm{GeV}\). The minimum lies at negative vacuum energy.}
    \label{fig:heavyH}
\end{figure}

Conversely,  Figure~\ref{fig:heavyH} displays the potential obtained for a heavy Higgs mass \(M_H = 400\,\mathrm{GeV}\) calculated at \(\mu = 320\,\mathrm{GeV}\). The potential becomes much steeper, and the minimum shifts towards lower compactification radii. However, the most important feature is that the effective potential is negative at the minimum. This behaviour illustrates the significant way a heavier scalar mass affects the vacuum structure, leading to a qualitative change in the nature of the stabilized compactification.

\medskip

The outcomes illustrate the sensitivity of the effective potential to the Higgs mass parameters. For small values of Higgs mass, one obtains a shallow potential with minimum at a positive energy ensuring a stabilized  and phenomenologically viable compactification. For larger Higgs mass values, the potential grows in magnitude and curvature.resulting in minima at negative vacuum energy and indicative of a qualitative change in the vacuum structure.

 \subsection{The non-SUSY AdS instability and implications for the 2HDM}

The swampland programme engenders an important expectation which is quantum gravity forbids completely stable non-supersymmetric anti-de Sitter vacua \cite{OoguriVafa2016,Vafa:2005ui,Palti:2019pca}. This expectation has been realized in several constructions in lower dimensions, such as the almost topological gravities in three dimensions, which can realize generically non-supersymmetric AdS solutions with either instabilities or decay channels consistent with the conjecture \cite{AlvarezGarcia2022}. More specifically, these non-SUSY AdS solutions see their instabilities manifest as either perturbative tachyonic fluctuations or by way of non-perturbative processes associated with an Euclidean tunnelling process, but ultimately granting that any non-SUSY AdS vacuum will see a finite probability of decaying.

Considering the Weak Gravity Conjecture (WGC), these instabilities have a natural interpretation, An AdS vacuum compatible with quantum gravity must possess a decay channel to charged or scalar states, preventing the existence of eternally stable configurations \cite{Ibanez2017,Ooguri:2006in}. The presence of decay modes is not coincidental but is a feature of any effective theory that can be UV-completed. On the other hand, the appearance of an AdS vacuum that seems stable and has no physical decay corresponding to the gravitational sector, should be taken to suggest inconsistency and indicate that we are in the swampland.
Before applying this reasoning to the compactified 2HDM, it is important to separate the local EFT analysis from possible microscopic decay channels. In this work, we only ask whether the three-dimensional radion--Higgs effective potential develops a locally stabilized non-supersymmetric AdS$_3$ minimum. Other non-perturbative instabilities can arise only after a microscopic stabilization mechanism is specified. For example, Kaluza--Klein vacua may decay through bubble-of-nothing processes, while flux-supported vacua can involve gravitational tunnelling or brane/membrane nucleation channels \cite{Witten:1981gj,Coleman:1980aw,Brown:1988kg,Bousso:2000xa}. Since the radion stabilization used here is introduced as a low-energy effective ansatz, rather than as a concrete flux- or brane-generated construction, these UV-dependent channels are not part of the present scalar-sector EFT. If such channels appear in a UV completion, they would simply provide additional decay mechanisms, and would therefore support rather than weaken the non-SUSY AdS interpretation adopted here.

When this reasoning is applied to the compactified Two-Higgs-Doublet Model (2HDM),  the results become especially important. The expanded scalar spectrum modifies the potential and the radion--Higgs interactions, both of which determine whether there is an AdS-like extremum that can move to a higher energy configuration. Radiative and loop-induced corrections in the 2HDM potential \cite{Coleman:1973jx,Ferreira:2019wjz,Cao:2023npo} can generate metastable regions, but to remain consistent with the AdS non-SUSY conjecture, the minima in these regions must remain unstable. We do not favor parameter spaces that provide local deep AdS minima without decay mechanisms available, but we remain compatible with swampland expectations \cite{Gonzalo2021,Nam2023,Casas2024} in those parameter spaces where instabilities may in fact be induced  by the extended Higgs/radion sectors themselves.

To clarify how the Higgs-sector parameters influence the vacuum structure, we present two additional examples of the effective potential. These illustrate how different Higgs masses lead to qualitatively distinct vacuum behaviours, thereby showing which compactified configurations remain stable and which tend toward instability.
    \begin{figure}[htbp]
    \centering
\includegraphics[width=0.75\linewidth]{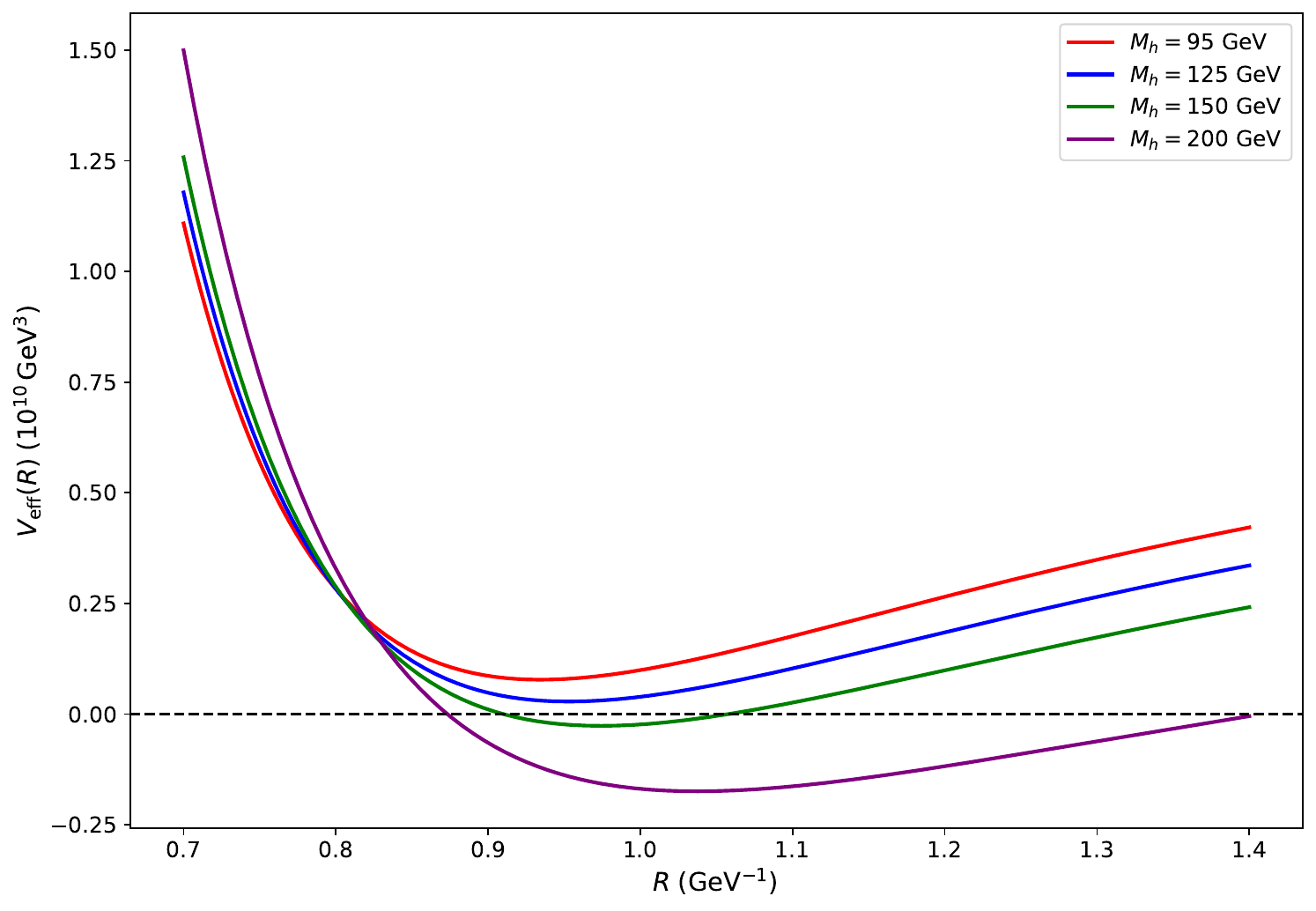}
    \caption{
    Three-dimensional effective potential \(V_{\rm eff}^{(3D)}(R)\) for
representative light-Higgs mass inputs. The radion and Higgs backgrounds are
kept fixed, and each curve corresponds to a different value of \(m_h\). }
    \label{fig:Veff-light-Higgs}
\end{figure}

\begin{figure}[H]
    \centering
\includegraphics[width=0.75\linewidth]{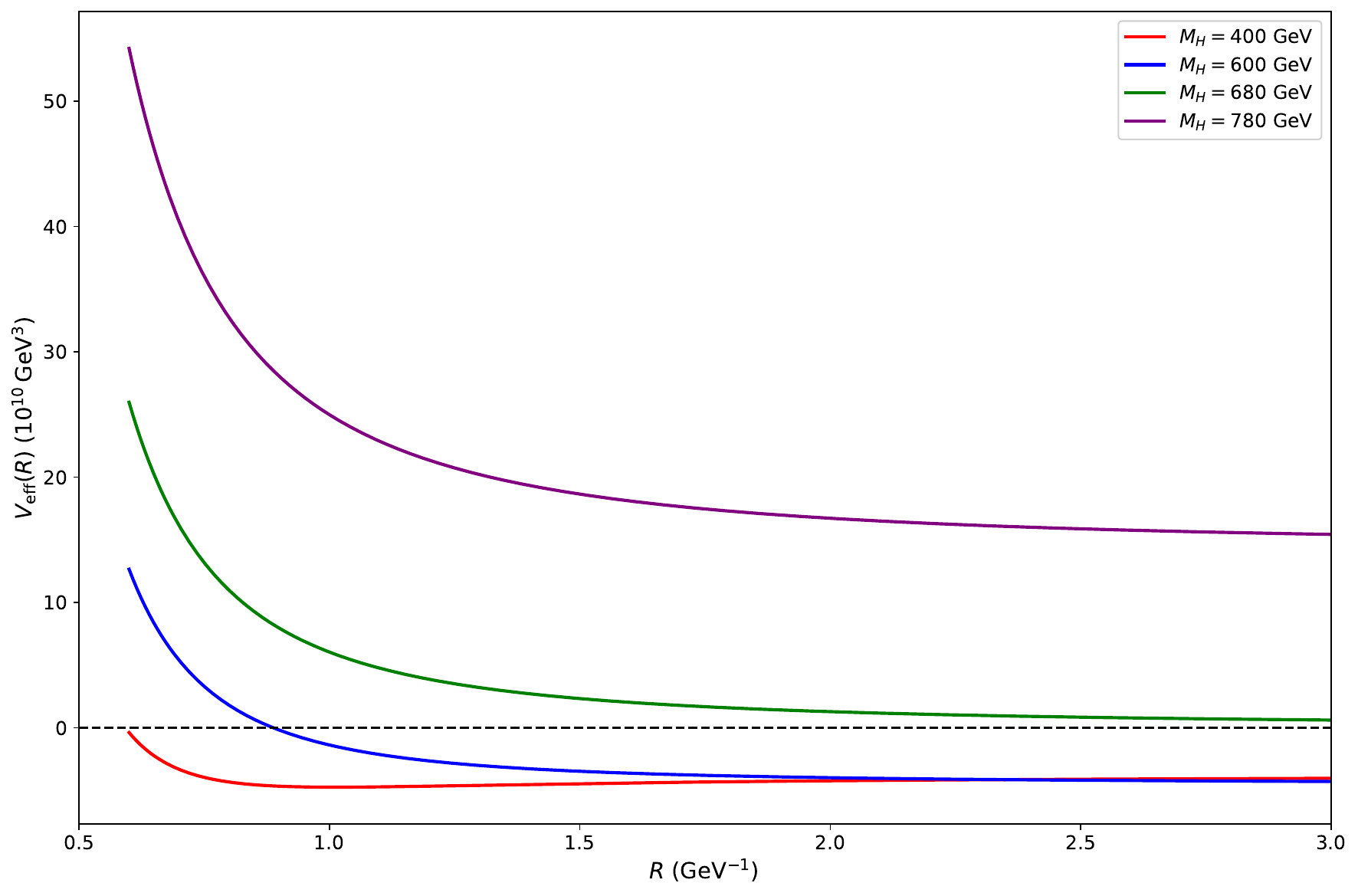}
    \caption{Three-dimensional effective potential \(V_{\rm eff}^{(3D)}(R)\) for
representative heavy-Higgs mass inputs. The radion and Higgs backgrounds are
kept fixed, and each curve corresponds to a different value of \(M_H\).}
    \label{fig:extraHeavy}
\end{figure}

\medskip

The first figure shows that light Higgs masses \(125\,\mathrm{GeV}\) lead to a stable and phenomenologically acceptable compactified vacuum, the effective potential shows a clear minimum with positive vacuum energy, which excludes any AdS-like behavior in the light-Higgs regime. These results are entirely consistent with what is expected from electroweak phenomenology, and shows that the radion stabilization mechanism naturally accommodates the observed Higgs mass and does not result in vacua instabilities.

The second figure illustrates the qualitative change as the scalar mass becomes heavy. As the mass increases, the location and depth of the minimum change, and for heavy Higgs masses less than roughly \(680\,\mathrm{GeV}\) the vacuum energy at the minimum goes negative. It is only for \(M_H \gtrsim 680\,\mathrm{GeV}\) that there is a return to a positive-energy minimum of the potential, which represents a physically viable compactified vacuum. When the non-supersymmetric AdS instability criterion is incorporated, this behaviour maps directly to a lower bound of about \(680\,\mathrm{GeV}\) on the Higgs mass, lighter values will produce an AdS-type minimum, while more massive values will embrace compatibility with a stable non-AdS compactified configuration.

Over the past few years, a number of both theoretical and
experimental investigations have explored the possibility of a
resonant structure in the $600\!-\!700$\thinspace GeV mass range. Our
radion--2HDM interpretation at $m_{H}\simeq 680$\thinspace GeV can be placed
in context by comparing it with alternative explanations proposed in the
literature. For instance, studies in the framework of type-II
2HDM have demonstrated that the $\sim 650$\thinspace GeV anomaly 
together with the $\sim 95$\thinspace GeV excess can be consistently 
accomodated within an enlarged scalar sector featuring suitable
mixing and decay modes~\cite{Benbrik:2025N2HDM}.\newline
Other works interprets the 650\thinspace GeV peak as a pseudoscalar
in three--Higgs--doublet models (3HDM) that manifest usually through a decay
chain of the form: $A\rightarrow h_{125}Z\rightarrow (\gamma \gamma
)(b\bar{b})$~\cite{Hmissou:2025THDM3}. Related studies on photon--initiated
Higgs--pair production processes in multi-doublet models have show
that heavy scalars up to 600-800\thinspace GeV could modify the di-higgs
signals, both at the LHC and at a future $\gamma \gamma $ collider~\cite%
{Arhrib:2025I1p2HDM}.

Moreover, direct experimental limits arise mainly from the CMS
search for new resonances that decay into two spin-0 bosons in the $%
\gamma \gamma b\bar{b}$ final state at $\sqrt{s}=13$ TeV~\cite{CMS:2023ggbb}%
, which gives the most stringent limits on $\sigma (pp\rightarrow
X)\times \mathrm{BR}(X\rightarrow \gamma \gamma b\bar{b})$ for $M_{X}\simeq
650\!-\!700$\thinspace GeV. Phenomenological analyses of the full CMS
Run2 datatset have also pointed out mild excesses in this mass window, 
further supporting the exploration of heavy scalar interpretation
\cite{Cea:2022QuantumLiquid}. And in contrast to N2HDM and 3HDM 
scenarios, the radion within our setup exhibits a distinct
coupling structure since its interactions, originating from the trace
of energy--momentum tensor, naturally enhance its gluon fusion
production with respect to QCD trace anomaly and modifies branching
ratios compared to purely doublet-like scalar.\newline
Consequently, the phenomenological acceptability of our $680$%
\thinspace GeV benchmark is largely determined by the expected radion
production cross section at 13\thinspace TeV and its branching ratios to
experimentally relevant channels like $\gamma \gamma b\bar{b}$ as well as $%
hh $ in conjunction with consistency with upper bounds posted by CMS. A
quantitative comparison of $\sigma (pp\rightarrow X)\times \mathrm{BR}%
(X\rightarrow \gamma \gamma b\bar{b})$ and $\sigma (pp\rightarrow X)\times
\mathrm{BR}(X\rightarrow hh\rightarrow b\bar{b}\gamma \gamma )$ with
the exclusion curves from the CMS 2HDM searches is therefore necessary to
assess whether a $680$\thinspace GeV resonance can arise from
radion--2HDM interplay.

\section{Discussion and Conclusion}

\label{sec:Discussion}

In this analysis, we have explored how a compactified Two-Higgs-Doublet Model
(2HDM), coupled to four-dimensional gravity on a circle \(S^1\), yields an
effective three-dimensional theory incorporating the compactified 2HDM scalar
sector, a radion degree of freedom, scalar Kaluza--Klein corrections, the
contribution inherited from the four-dimensional vacuum energy, and an effective
radion-stabilizing potential. The latter is treated as a low-energy ansatz
motivated by the existing stabilization literature describing the general
behavior of modulus-stabilized fields.

The analysis using radiative corrections reveals that the compactification
radius \(R\) is not just an externally fixed background dimension; rather,
\(R\) can be dynamically selected through the interplay among the radion
potential, the Higgs-sector contribution to the vacuum energy, and the scalar
Kaluza--Klein corrections. For a Standard Model-like Higgs mass of
\(m_h \approx 125\,\mathrm{GeV}\), this leads to an effectively stabilized
compactification in the effective theory considered here.

The incorporation of the heavier scalar states associated with the 2HDM modifies
the depth and location of the compactified minimum as a function of the heavy
Higgs mass. Specifically, for scalar masses below a critical value,
\(M_H \simeq 680\,\mathrm{GeV}\), we find that the effective potential tends to
possess negative-energy minima. However, for scalar masses above this critical
value, the effective potential returns to a non-AdS-like minimum. Thus, we
interpret \(M_H \simeq 680\,\mathrm{GeV}\) as a limiting value in our numerical
analysis of the compactified effective field theory, not as a universally
applicable, model-independent bound.

Based on the non-supersymmetric AdS instability conjecture, we also interpret
perturbatively stable non-supersymmetric AdS\(_3\) minima as
swampland-disfavored regions of parameter space. Hence, the sign of the
three-dimensional vacuum energy associated with the radion-stabilized minimum
is an important diagnostic tool for examining whether the compactified 2HDM
generates vacua consistent with quantum-gravity expectations. Our interpretation
suggests that, within the simulations performed here, the heavy Higgs sector is
constrained to lie above approximately the \(680\,\mathrm{GeV}\) scale in order
to produce non-AdS-like compactified vacua.

The compactified Two-Higgs-Doublet Model provides a clear framework for studying
how extra-dimensional dynamics, radion stabilization, and extended Higgs-sector
phenomenology interact. The results indicate that Swampland-inspired consistency
requirements can offer interesting constraints on scalar sectors beyond the
Standard Model when the lower-dimensional vacuum structure of the theory is
taken into account. A more complete analysis could include the full light-particle
spectrum, gauge and fermionic contributions, and additional Casimir terms, thus
offering further refinements of the radion dynamics and the resulting
phenomenological constraints.

\section*{Acknowledgments}
The work of M. A. Rbah, S. Saoud and R. Sammani is funded by the National Center
for Scientific and Technical Research (CNRST) under the PhD-ASsociate Scholarship
(PASS).
\bibliographystyle{unsrt}
\bibliography{references}

\end{document}